\documentclass[aps,prl,twocolumn,showpacs]{revtex4}  
\usepackage{epsfig}
\begin{document}
 \title{Unitarity of scattering and edge spin accumulation} 
 \author{Alexander Khaetskii$^{1}$ \thanks{On leave from Institute of Microelectronics Technology, Russian Acedemy of Sciences, 142432 Chernogolovka, Moscow District, Russia}, 
and Eugene Sukhorukov$^2$}
\affiliation{$^1$  Department of Physics, University at Buffalo, SUNY, Buffalo, NY 14260-1500
} 
\affiliation{$^2$ Department of Theoretical Physics, University of Geneva,
24 quai Ernest Ansermet, CH-1211, Switzerland}

\date{\today}


\begin{abstract}
We consider a 2D ballistic  and quasi-ballistic structures with spin-orbit-related  splitting of the electron spectrum. The ballistic region is attached to the leads with a voltage applied between them.
 We calculate the edge spin density which arises in the presence of a charge current through the structure.
We solve the problem  with the use of the method of scattering states and clarify the important role of the unitarity of scattering. 
In the case of a straight boundary  it leads to exact cancellation of  long-wavelength oscillations of the spin density.  
 In general, however, the smooth spin oscillations with the spin precession length may arise, as it happens, e.g., for the wiggly boundary. 
Moreover, we show that the result  crucially depends on the form of the spin-orbit Hamiltonian.

 \end{abstract}
\pacs{72.25.-b, 73.23.-b, 73.50.Bk}

\maketitle

   Currently,  there is a great interest, both experimental and theoretical, to spin currents and spin accumulation in various mesoscopic semiconductor structures \cite{Rashba,We}. Both phenomena  are due to spin-orbit  (s-o) coupling and are  of great importance for the future of spin electronics. 
The edge electron spin density accumulation, related to the Mott asymmetry in electron scattering  off impurities,  has been recently measured \cite{Kato}.  
  Moreover, the edge spin density in the  two-dimensional (2D)  hole system, which is due to the intrinsic mechanism \cite{Sinova} of the s-o interaction,  has also been observed \cite{Wunderlich}.  
   It is well known \cite{We}, that in the diffusive regime (and when a spin diffusion length is much larger than a mean free path), 
the spin density appearing near the boundary is entirely determined by the spin flux coming from the bulk.  For example, in the diffusive regime and in the case of the Rashba Hamiltonian, when the spin current in the bulk is zero, the spin density component perpendicular to the plane is zero everywhere down to the sample boundary \cite{BC}. 
   \par   In an opposite case, when the spin precession length is much shorter than the mean free path, the situation is much less investigated.  
  An example of such a system is a mesoscopic structure with s-o-related splitting of the electron spectrum  $\Delta_R$, in the limit $\Delta_R \tau_p \gg 1$, where $\tau_p$ is the mean free time. Besides, in the presence of s-o interaction, the boundary scattering itself is the source of generation of the spin density. It is obvious, that the characteristic length near a boundary,  where the spin density arises is the spin precession length, $L_s=\hbar v_F/\Delta_R$, with $v_F$ being the Fermi velocity. 
  This mechanism of the spin density generation is the subject of our paper. 
We show that various situations may arise, depending on the form of the  s-o Hamiltonians.

\par
\begin{figure}
\begin{center}
\epsfig{file=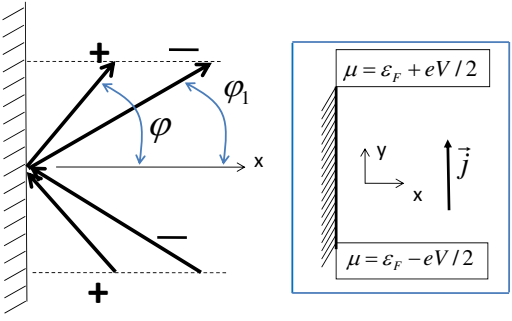,width=0.45\textwidth}
\end{center}
\caption{Left: Schematics of the boundary specular scattering in the presence of spin-orbit coupling. Plus and minus modes are shown for the same energy and the same wave vectors along the boundary. Right: Geometry of the system. Voltage $V$ applied to the ideal leads causes a charge current through the ballistic region.}
\label{fig:Spin1}
\end{figure}

  We  start with a 2D system described by the Rashba Hamiltonian  in 
the {\it ballistic} limit,  where a mean free path  is much larger than the sample sizes. The ballistic  region is attached to the leads, and a voltage $V$ applied between the leads  causes a charge current through the structure, as shown in Fig.\ \ref{fig:Spin1}.  
Since the electric field is absent inside an ideal ballistic conductor, the edge spin polarization appears not as a result of the acceleration of  electrons by an electric field, but 
rather due to the difference in populations of left-moving and right-moving electrons.   
The combined effect of boundary scattering and spin precession leads to oscillations of the edge spin polarization.
 Note that there is no relation between the spin current in the bulk, which is zero in the considered case, and the edge spin accumulation.  

 \par
 The problem of the spin density accumulation in a ballistic system and for a straight boundary has been considered  earlier in Refs.\ [\onlinecite{Zyuzin},\onlinecite{Reynoso}] 
with the help of the Green's functions method.  
Surprisingly, in this case the long-wavelength oscillations of the spin density cancel,  and the final result contains only  Friedel-like oscillations with  the momentum $2k_F$. 
 This effect may be interpreted as s-o splitting of the Friedel oscillations in the charge density: two charge oscillations corresponding to spin-up and spin-down orientations get shifted with respect to each other in the presence of the s-o interaction. 
Therefore, strictly speaking,  this phenomenon is different from a s-o-related  accumulation of the spin density upon boundary scattering.
 Besides, the method used in Ref.\ [\onlinecite{Zyuzin}] does not allow to understand the reason for the cancellation of  long-wavelength  oscillations of the spin density.  
\par
 We solve the problem of edge spin accumulation by using scattering theory, with scattering states coming from different leads of the structure and, therefore, having different occupations. The simplicity of the method allows us to gain an insight into the underlaying physics.  We show that it is the unitarity of  scattering that leads to the exact cancellation of long-wavelength  oscillations  of the spin density with the period  $L_s$ in the case of a straight boundary. It should be also mentioned that the observed behaviour is closely related to the effective one-dimensional character of scattering,  arising from the translational invariance along the boundary. 
However, the case of a straight boundary appears to be a rather exceptional one.
 In general,  smooth spin oscillations with the spin precession length  $L_s$ arise, as it happens, for example, for the wiggly boundary, or for scattering off a circular impurity in a 2D electron system \cite{2Dcase,Khaetskii1}. This is a consequence of the fact that in higher dimensions the conditions of the unitarity of scattering take a different form, as explained below.  In all these situations, the spin density decays as a power law of the distance from the scatterer.

    \par 
       Rashba s-o Hamiltonian  in the bulk of a ballistic 2D electron system takes the following form
\begin{equation}
\hat {\cal H}({\bf p})=\frac{p^2}{2m}+ \frac{\alpha}{2}\vec{n}[\vec{\sigma}\times {\bf p}],
\label{Rashba}
\end{equation}
where $\vec{n}$ is the normal to the plane, $\vec{\sigma}$ are the Pauli matrices, and ${\bf p}$ is the 2D momentum. 
The solutions of this Hamiltonian corresponding to  the helicity values $ M=\pm $ have the form $\exp (i{\bf p}{\bf r}/\hbar)\chi_M({\bf p})$, where ${\bf r}=x,y$.  
The explicite form of the spinors and their eigenenergies is
$$\chi_{\pm}(\varphi)= \frac{1}{\sqrt{2}}\left( 
\begin{array}{ll}
1 & \\
\mp i e^{i\varphi} &
\end{array}
\right),  \,\,\,  \epsilon_M(p)= \frac{p^2}{2m}+ \frac{M}{2}\alpha p,$$
with $\varphi$ being the angle between the momentum ${\bf p}$ and the positive direction of the $x$-axis. 
\par
We consider the semi-infinite system and choose the $x$-axis to be directed perpendicular to the boundary ($x=0$) of the 2D system  (see Fig.\ \ref{fig:Spin1}). 
The wave functions, which obey zero boundary conditions 
at $x=0$,  are obviously the {\it scattering states}, which constitute the complete set of the orthonormal functions.  
Two scattering states corresponding to incident plus and minus modes with given wave vector along the boundary and the same energy are
\widetext
\begin{eqnarray}
\hat{\Psi}_{+}^{(0)}(x,y)=e^{ik_yy}[\chi_{+}(\pi -\varphi)e^{-ikx}+ F_{+}^{+}\chi_{+}(\varphi)e^{ikx}+               F_{+}^{-}\chi_{-}(\varphi_1)e^{ik_1x}]; \,\,\,  \hat{\Psi}_{+}^{(0)}(0,y)=0, 
\label{plusmode} \\
\hat{\Psi}_{-}^{(0)}(x,y)=e^{ik_yy}[\chi_{-}(\pi -\varphi_1)e^{-ik_1x}+ F_{-}^{+}\chi_{+}(\varphi)e^{ikx}+               F_{-}^{-}\chi_{-}(\varphi_1)e^{ik_1x}]; \,\,\,   \hat{\Psi}_{-}^{(0)}(0,y)=0.
\label{minusmode}
\end{eqnarray}
\endwidetext
 Here, the wave vectors are defined as follows
 \begin{equation}
 k^2=k_+^2-k_y^2, \,\,\, k_1^2=k_-^2-k_y^2, \,\,\, \hbar k_{\pm}=m(v_F \mp \frac{\alpha}{2}),
 \label{k_vectors}
 \end{equation}
 where $p_{\pm}=\hbar k_{\pm}$ are the momenta at the Fermi energy in the plus and minus modes.  The angles $\varphi$, $\varphi_1$ may be expressed as $\sin (\varphi)=k_y/k_+$ and $\sin (\varphi_1)= k_y/k_-$ (see Fig.\ \ref{fig:Spin1}).  
\par
 From Eqs.\ (\ref{plusmode}) and (\ref{minusmode}), one finds the scattering amplitudes $F_{+}^{+}$ and  $F_{+}^{-}$:
\begin{equation}
F_{+}^{+}=-\frac{(e^{i\varphi_1}-e^{-i\varphi})}{(e^{i\varphi_1}+e^{i\varphi})}; \,\,\, F_{+}^{-}= -\frac{2\cos\varphi}{(e^{i\varphi_1}+e^{i\varphi})}.
\label{ScattAmpl}
\end{equation}
 One can check that the amplitudes $F_{-}^{-}$ and  $F_{-}^{+}$ for the incident minus mode with the same $k_y$ and the same energy  are obtained from $F_{+}^{+}$ and  $F_{+}^{-}$ by replacing  $\varphi \leftrightarrow  \varphi_1$. Then, the components of the unitary scattering matrix $\hat{S}$ acquire the following form:
  \begin{equation}
  S_+^+=F_{+}^{+},\,\, S_-^-=F_{-}^{-}, \,\, S_+^-= S_-^+=F_{+}^{-}\sqrt{\frac{v_{x,-}}{v_{x,+}}},
  \label{SMatrix}
  \end{equation}
 where $v_{x,i}=\partial \epsilon_i/\partial p_x$ are the group velocities. For the Rashba model one has 
 $v_{x,-}/v_{x,+}= \cos\varphi_1/\cos\varphi$. 
 \par 
   The wave functions Eqs.\ (\ref{plusmode}) and (\ref{minusmode}) may now be used to calculate the average $z$-component of the spin as a function of coordinates:
  \begin{eqnarray}
  \langle S_z(x)\rangle =\sum_{i=\pm}\int \frac{dk_y}{(2\pi)^2}\frac{d\epsilon}{v_{x,i}}f_F(\epsilon,k_y) \nonumber \\
 \times  \langle\hat{\Psi}_{i}^{(0)}(x)|\hat S_z |\hat{\Psi}_{i}^{(0)}(x)\rangle,
  \label{S_z}
\end{eqnarray}
 where $f_F(\epsilon,k_y)$ is the Fermi distribution function, which takes either of two values,  
  $f_F(\epsilon -\mu -eV/2)$ or $f_F(\epsilon -\mu +eV/2)$,  depending on the sign of $k_y$. 
We find,  that one may distinguish various contributions to $\langle S_z(x)\rangle $ with different oscillation periods, which originate from an interference of different terms in  Eqs. (\ref{plusmode}) and (\ref{minusmode}). 
  The smooth part of $\langle S_z(x)\rangle_s $,  which involves the interference of the outgoing waves [two last terms in Eqs.\ (\ref{plusmode}) and (\ref{minusmode})], reads:
  \begin{eqnarray}
\langle S_z(x) \rangle_s \propto  \int dk_y d\epsilon f_F(\epsilon,k_y) \frac{1}{\sqrt{v_{x,-}v_{x,+}}} \nonumber \\ 
\times
  \Bigg [A\langle \chi_{-}(\varphi_1)|\hat{S}_z|\chi_{+}(\varphi)\rangle e^{i(k-k_1)x} +{\rm c.c.}\Bigg ],
    \label{Unitary}
\end{eqnarray}
where 
$$
 A=S_{+}^{+}\cdot (S_{+}^{-})^{\star} + S_{-}^{+} \cdot (S_{-}^{-})^{\star}.
 $$
Here we used the fact that the distribution function $f_F(\epsilon,k_y)$,  describing a particular lead,  has the same value at given energy 
  for  plus and minus mode.    Note, that the period of oscillations of the exponential factor $e^{i(k-k_1)x}$ in Eq.\ (\ref{Unitary}) is of the order of the spin precession length. 
   However, the term (\ref{Unitary}) vanishes, because the expression $A$ is nothing but a non-diagonal component of the identity matrix $ \hat{S}\hat{S}^{\dagger} $. 
   Thus, we obtain an interesting result,  that the only reason for the cancellation of the long-wave length oscillations with the period $L_s$  in  $\langle S_z(x)\rangle $ is the unitarity of scattering.
          \par
  By taking  into account in Eq.\ (\ref{S_z}) the terms responsible for the interference between incoming and the outgoing waves [for example, between first and second terms in Eq.\ (\ref{plusmode})], and adding the contribution from the evanescent modes \cite{evanes}, we reproduce Eq.\ (16) of the Ref.\ \cite{Zyuzin}.  It can be written in the form $\langle S_z(x)\rangle= (eV/8\pi^2mv_F^2){\rm Im} I $, where
$$
I=\int _{0}^{k_-}dk_y\frac{k_+k_-  + k_y^2 -kk_1}{k_y}(e^{ikx}-e^{ik_1x})^2.
$$
Note, that in the interval $k_+<k_y<k_-$,  the quantity $k$ has purely imaginary value, which corresponds to the evanescent modes. From this form of the presentation, we can immediately see that  $\langle S_z(x)\rangle $ contains  only $2k_F$ component,  while all the long-wavelength  oscillations cancel exactly.  
\begin{figure}
\begin{center}
\epsfig{file=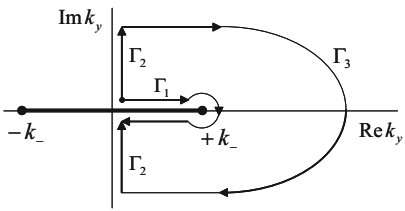,width=0.3\textwidth}
\end{center}
\caption{The original contour $\Gamma_1$ along the real axis can be deformed into the part  $\Gamma_2$ going along the imaginary axis, and the part $\Gamma_3$ going far from the origin.}
\label{fig:Spin2}
\end{figure}
Indeed, with the branch cut along the real $k_y$ axis  between the points $-k_-$ and $+k_-$  (see Fig.\ \ref{fig:Spin2}),  the integrand function in $I$ is an analytical function of the variable $k_y$ in the right half plane ${\rm Re} k_y> 0$ (for positive $x$). Since we need the imaginary part of $I$, the integration is going alone the upper edge of the branch cut from 0 up to $+k_-$ and then back along the lower edge of the branch cut. Because of the analyticity mentioned above, this integral is equal to the one taken along the imaginary axis of $k_y=i\kappa$.  Then, for $x\gg \lambda_F$  the latter integral is determined by small $\kappa\ll k_F$:
$$
I\simeq -2 (e^{ik_-x} -e^ {ik_+x})^2 \int_0^{\infty}d\kappa \kappa e^{ix\kappa^2/k_F}, 
$$
which gives for the spin density $\langle S_z(x)\rangle  \approx (eV/2\pi^2 v_F x)  \cos(2mv_Fx)\sin^2(\alpha m x/2)$, coinciding with the result of Ref. \cite{Zyuzin}.  Therefore, the total spin per unit length along the boundary scales as $\int_0^{\infty}dx \langle S_z(x)\rangle  \propto \alpha^2$. Note, that the main contribution to this integral comes from small distances from the boundary, $x\simeq\lambda_F$.
 
        \par The cancellation of smooth spin density oscillations in case of the Rashba Hamiltonian and straight boundary occurs also in the quasi-ballistic situation:  $L\gg l\gg L_s$,  where $L$ is the sample size, and $l$ is the mean free path.   In this case, the electric field  in the bulk of the sample is finite. Therefore,  the distribution functions for the plus and minus modes,  $f_{++}(\vec{k}_+)$ and  $f_{--}(\vec{k}_-)$,  are determined by the electric field and by scattering off  the impurities in the bulk of a system \cite{Khaetskii}.  The wave vectors $\vec{k}_+$ and  $\vec{k}_-$, shown in  Fig.\ \ref{fig:Spin1},  correspond to a given energy and a given wave vector along the boundary.
 In the quasi-ballistic case considered here, these functions are equal, i.e.  $f_{++}(\vec{k}_+) = f_{--}(\vec{k}_-)$, similar to a ballistic situation.  Under such a condition, the unitarity of scattering, see Eq.(\ref{Unitary}), leads to the cancellation of smooth edge spin density oscillations, in contrast to what has been stated in the literature \cite{Sonin}. 
 Indeed,  when the electric field is parallel to the boundary, the distribution functions in questions are  $f_{++}(\vec{k}_+) = f_{++}(k_+) \sin\varphi $, and $f_{--}(\vec{k}_-) = f_{--}(k_-) \sin\varphi_1$ (see Eq.\ (9) of Ref.\ \cite{Khaetskii}).  For the case of the Rashba Hamiltonian, the following relation has been  obtained $k_+ f_{--}(k_-)=k_- f_{++}(k_+)$ \cite{Note}.  Then,  the ratio  is $f_{++}(\vec{k}_+) /f_{--}(\vec{k}_-) =k_+\sin\varphi/k_-\sin\varphi_1=k_y/k_y=1$.           
\par 
Depending on the form of the s-o Hamiltonian, the unitarity may show up in totally different ways,  leading, in general,  to different patterns of the edge spin density. 
In particular,   boundary conditions play a crucial role. 
 Let us consider 2D holes, described by the cubic (in 2D momentum) s-o Hamiltonian, still keeping in mind the ballistic case and an abrupt straight boundary. 
The unitarity condition, i.e. the charge flux conservation,  acquires a different form. This may be seen already for the case of a normal incidence, where  the Hamiltonian  has the form $p_x^2/2m + \alpha \hat {\sigma}_x p_x^3$.  In the case of the plus incident mode, applying previous zero boundary conditions (which sometimes are used for this Hamiltonian in a numerical treatment of the edge spin accumulation problem),  one obtains  $F_+^+=0$, $F_+^- =i$. In other words,  after elastic backscattering the helicity value changes  sign, which is a consequence of the fact that ${\sigma}_x$ is conserved.  Then,  for the  charge flux to be conserved,  one needs the equality of the group velocities $v_+$ and $v_-$, corresponding to the plus and minus modes at the same energy.  
However, in contrast to the case of the Rashba Hamiltonian, where  $v_+=v_-$ ,  those velocities are not equal for the cubic Hamiltonian:  $v_+-v_-=\alpha (p_++p_-)^2$.   
The formal way to resolve the trouble is to note that the cubic Hamiltonian  has three solutions for a given energy, one of them corresponding to a momentum larger than $k_F$  (for small spin-orbit coupling).  Thus, in general,  the unitarity of scattering in slow  plus and minus channels is violated,  because some flux is carried away by a fast oscillating mode. Therefore, the corresponding coefficient  $A$ in Eq.\ (\ref{Unitary}) does not vanish, and smooth spin density oscillations occur. 
However,  for a physically relevant confining potential which grows to infinity in a smooth manner,  the fast oscillating modes are not excited. Then,  the unitarity of scattering into slow  modes is restored,  and the smooth spin density oscillations do not appear.  

\begin{figure}
\begin{center}
\epsfig{file=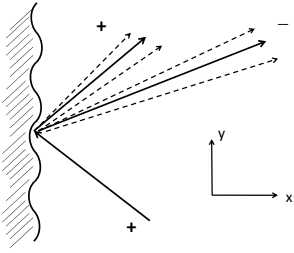,width=0.27\textwidth}
\end{center}
\caption{Schematics of scattering of the plus incident mode by a wiggly boundary, $x=W\sin(2\pi y/\lambda)$. Apart from the main scattering channels (solid lines), there are additional scattering waves with the wave vectors along the boundary shifted by $\pm 2\pi/\lambda$ (dashed lines).}
\label{fig:Spin3}
\end{figure}

Let us consider  scattering off a wiggly boundary, shown in Fig.\ \ref{fig:Spin3},  for the case of  the Rashba Hamiltonian.  In this case the translational invariance is broken, therefore the condition of the unitarity of scattering takes a different form,  as compared to the case of a straight boundary.  As a result, the cancellation of the smooth spin density oscillations does not take place,   leading to the total spin that is not small in the parameter $\alpha$ \cite{Silvestrov,Silvestrov1}.  In order to demonstrate this effect, we consider the mathematically simple case of the abrupt impenetrable boundary described by the equation: $x=\zeta(y)\equiv W\sin (2\pi y/\lambda)$. To the lowest order in $W$,  the boundary condition reads:
      \begin{equation}
     \hat{\Psi}(0,y)+ \zeta(y)\frac{d\hat{\Psi}(0,y)}{dx}=0
     \label{bouncond}
     \end{equation}
   We are looking for the solution in the perturbative form $ \hat{\Psi}_{\pm}(x,y)= \hat{\Psi}_{\pm}^{(0)}(x,y)+ \hat{\Psi}_{\pm}^{(1)}(x,y)$, where  the zeroth order functions are given by Eqs.\ (\ref{plusmode}) and (\ref{minusmode}).  The first order correction, proportional to $W$,  is the 
   superposition  of scattering waves with the wave vectors along the boundary shifted by $\pm 2\pi/\lambda$ (see  Fig.\ \ref{fig:Spin3}),   and with the $k$-vectors in the x-direction given by: 
    $$
    k^{\frac{>}{<}}=\sqrt{k_+^2-(k_y \pm \frac{2\pi}{\lambda})^2}, \,\,\,  k_1^{\frac{>}{<}}=\sqrt{k_-^2-(k_y \pm \frac{2\pi}{\lambda})^2}. 
    $$
     
    \par
    From now on,  we assume  $\lambda_F \ll L_s \ll \lambda$.  In addition,  in order to obtain an analytical expression for the spin density, we consider the case $x/ \sqrt{\lambda \lambda_F} \ll 1$. 
 In contrast to the case of the straight boundary,  there are oscillations with three different periods: $2k_F$- oscillations,  and  the oscillations with two long periods,  $\xi$ and $L_{s}$. Here $\xi=1/\sqrt{k_-^2-k_+^2}$ is the new length scale. Under the conditions considered in the paper we obtain the set of inequalities $\lambda_F \ll \xi \ll L_{s}$, where $L_{s}= 1/(k_- - k_+)=\hbar/m\alpha$ is the spin precession length. 
 If $k_y \rightarrow k_+$ (i.e.,  $k \rightarrow 0$), then $k_1$ tends to $1/\xi$, which clarifies the physical meaning of $\xi$. 
  
  \par
For the contribution of the  long wavelength oscillations,  we obtain  \cite{WeJETPLett}:
\begin{eqnarray}
    \langle S_z(x,y)\rangle = \frac{eV}{(2\pi)^2\hbar v_F}(\frac{2\pi W}{\lambda})\cos(\frac{2\pi y}{\lambda})I_{long}(x), \nonumber \\
   I_{long}(x)= \frac{2\sin(\frac{x}{\xi})}{\xi}+\frac{2\cos(\frac{x}{\xi})}{x}+
  \frac{\pi}{2L_{s}}N_1(\frac{x}{L_s})-\frac{1}{x} \nonumber \\
+\frac{2x}{\xi}\frac{\partial}{\partial x} \int_0^{1}dze^{-(x/\xi) z} \cos\frac{x\sqrt{1-z^2}}{\xi}, 
\label{total}
\end{eqnarray}
where the last term is the contribution of the evanescent modes, and $N_1(x)$ is the Bessel function of the second kind. 
At the distances $x\ll \xi$,  we obtain the following dependence 
$I_{long}=-2x^2/3\xi^3 + (x/2L_{s}^2)[\gamma +\ln(x/2L_{s})]$. 
In the opposite limit, $x\gg \xi$, we  find
$I_{long}=\frac{\pi}{2L_{s}}N_1(x/L_{s})-\frac{1}{x}+ \frac{2\cos(x/\xi)}{x}$.  
At even larger distances, $x \gg L_{s}$, one obtains  smooth oscillations with the period of the order of $L_{s}$,  and the amplitude being proportional to $\sqrt{\alpha}$. 
Note,  that the total spin per unit length along the boundary is proportional to the integral $\int dx I_{long}(x)\simeq (\pi/2L_s)\int_{\xi}^{\infty}dx N_1(x/L_s)-\int_{\xi}^{\sqrt{\lambda \lambda_F}}dx/x \simeq -\ln(L_s/\xi)-\ln(\sqrt{\lambda \lambda_F}/\xi)\simeq -(1/2)\ln(\lambda/\lambda_F)$, i.e.,  it is not small in s-o coupling,  in contrast to the case of a straight boundary.

\par 
In conclusion, we have considered the problem of the edge spin accumulation in mesoscopic structures with spin-orbit-related splitting of the energy spectrum,  when the associated spin precession length is much smaller than the mean free path.  In the presence of the charge current,  the spin density develops oscillations near an edge in the direction transverse to the boundary. 
We have used the scattering states method  to clarify the physics associated with this effect. 
 The result  crucially depends on the form of s-o Hamiltonian and the boundary conditions. 
The unitarity of scattering  in the case of a straight boundary and Rashba Hamiltonian leads to the cancellation  of long-wave length spin density oscillations.
On the contrary, the spin density in the case of  wiggly boundary  oscillates with a large period of the order of the spin precession length, quite similar to the case  of electron scattering  off antidots in 2D system with the Rashba Hamiltonian \cite{Khaetskii1}. 

 \par 
 We acknowledge the financial support from the Program "Spintronics" of RAS, the Russian Foundation for Basic Research (Grant No. 07-02-00164-a), the Swiss National Science Foundation and SPINMET project (FP7-PEOPLE-2009-IRSES).


\begin{thebibliography}{99}
\bibitem{2Dcase}
 Only in the effective 1D situation the unitarity of scattering leads to the cancellation of long-wavelength oscillations of the spin density.  This is no longer the case for 2D systems. 
The  problem for electrons,  described by the Rashba Hamiltonian, scattered off a circular scatterer (an antidot) may be solved exactly in the presence of a charge current. 
The  $S_z$-component of the spin density, oscillating with period $L_s$, appears around the antidot  \cite{Khaetskii1}.
\bibitem{Silvestrov}
Note,  that  Ref.\ \cite{Silvestrov1} considers the case of a straight  boundary,  which is smooth  in the transverse direction,   instead of abrupt one.  
If the width of the  boundary  is larger than $L_s$,  two neighboring strips of the spin density of the opposite signs arise. 
We stress,  that the physics discussed in Ref. \cite{Silvestrov1} is different from that we consider here.  Since the boundary is still straight, 
the unitarity of scattering leads to the cancellation of smooth oscillations with the period $L_s$.    As a result,  the total spin, found in  \cite{Silvestrov1},  is small and proportional to $\alpha^2$.
\bibitem{evanes} When $k_+<k_y<k_-$,  
 the wave vector in x-direction is purely imaginary  for the reflected plus mode. This mode is called evanescent.  
This situation takes place only for the incident minus mode, since only this mode comes from the leads 
under the indicated  conditions.
\end{thebibliography}
\end{document}